# Conformational dynamics of charged polymers interacting with charged nanoparticles

Yuanzheng Zhu, Zixiang Liu, Michael T. Griffin, David N. Ku, and Cyrus K. Aidun

*G. W. Woodruff School of Mechanical Engineering at Georgia Institute of Technology, Atlanta, Georgia 30332, USA*

The transition from globular to elongated states of biopolymers in shear flow occurs at a distinct critical shear rate, $\dot{\gamma}^*$. The magnitude of $\dot{\gamma}^*$ depends on the internal potential and the polymer length. For example, the critical shear is much larger for von Willebrand Factor (vWF) compared to DNA. Furthermore, it is shown through computational analysis of vWF (model-vWF) that $\dot{\gamma}^* \sim N^{1/3}$, where $N$ is the number of dimeric units in the vWF. In this study, we show that in the presence of charged nanoparticles (CNP) and polymer, the critical shear rate scales differently depending on the polymer length and the charge-strength of the CNP. It is shown that CNP alter the conformational dynamics of polymers under shear flow when the polymer beads have an opposite charge. The introduction of CNP shifts the critical shear rate and alters the scaling. Furthermore, it is shown that the critical zeta potential of CNP, $\zeta^*$, scales linearly with shear rate, and scales cubically with ratio $\beta$ between final CNP-polymer composite size and original polymer size, that is $\zeta^* \sim \dot{\gamma}\beta^{-3}$.

The conformational dynamics of polymers suspended in viscous shear flow have been extensively studied over the past 60 years [1-5]. Polymers are often modeled as a chain of beads where neighboring beads are connected by a spring-like link. The equilibrium state in quiescent viscous fluid is a globular 'collapsed' state. Under shear stress the globular state loses stability and elongates at a critical flow parameter, depending on the nature of the flow. In simple shear flow, where there is a combination of extension/compression and rotation, the polymer undergoes complex conformational dynamics. The time-averaged extensional length increases with shear rate until at high shear where it reaches one-half of the contour length, on average [1]. In extensional flow, the polymer elongates due to continuous extension. The distortion of polymer coils from an equilibrium globular state to an elongated state and the average orientation have been measured by light scattering method under various flow conditions [6,7].

The conformational change of biopolymers in flow is of great importance, as it relates to the functionality of polymer molecules like DNA and von Willebrand factor (vWF). Studies of the shear-induced behavior of DNA [8,9] and vWF molecules [10-13] in bulk fluid show that these biopolymers oscillate under shear, exhibiting periodic elongation, relaxation and tumbling behavior. In the case of vWF, while in globular shape, most of the reactive domains are masked inside the globule, unavailable to interact or bind with other molecules. However, when in large enough shear or elongational flow, vWF unfolds and exposes binding domains. For example, interaction of platelet receptors (GPIba) to the binding domain (A1) on the vWF monomer in hemostasis and thrombosis occurs more readily when the vWF is elongated. In fact, adjusting levels of shear rate is a biological self-regulatory mechanism to control activity of the protein.

It is common to model polymers by beads attached by a spring [9,14-16]. For the case of vWF, each bead represents a dimer. Several studies of the conformation of model-vWF and actual vWF in shear flow show that the model can effectively capture the deformation of the vWF based on the flow environment [10,11,17].

Surface charge is an important feature in the interaction of biopolymers with cells. The interaction between vWF and platelets in hemostasis or thrombosis is based on opposite electrostatic surface charge of the A1 (+) domain of the vWF and the platelet GPIba (-) receptor [18]. It is shown that by introducing negatively charged nanoparticles (CNP), the time scale for experimental thrombosis will vary significantly [19]. This effect is attributed to the interaction of the CNP with the positive A1 domain of vWF. In the model-vWF, it is assumed that the whole dimer (i.e., each bead) has a net positive point-charge equal to twice the potential of the A1 domain, since there are two A1 domains in a dimer [20].

The critical (or threshold) shear rate, $\dot{\gamma}^*$, where transition of the polymer from only coiled to periodically stretched conformation occurs, depends on the number of beads (N), or length of the polymer, as well as other parameters. The relation between $\dot{\gamma}^*$ and $N$ depends on the polymer and the interaction potential between the beads. As previously reported, the magnitude of $\dot{\gamma}^*$ is shown to scale as $\dot{\gamma}^* \sim N^{1/3}$ in the case of model-vWF [10].

However, none of these models have included the electrostatic charge of the A1 domain. In this study, we explore the effect of CNP interaction with model-vWF. In particular, we show that there exists a unique scaling for the critical shear rate depending on charge strength of CNP and size of the polymer.

We first present the mathematical details of the model-vWF. To establish confidence in the model, the results are compared with available experimental data for the characteristic time scale for model-vWF relaxation in quiescent fluid and critical shear rate for vWF elongation [21,22]. We then include the surface charge (-) for the CNP and (+) for the model-vWF beads. The effect of the CNP on the conformational dynamics of the model-vWF and the modification of the scaling parameters will be presented.

**Model-vWF**

It is common to model polymers as beads attached with spring-like potential. The interaction of beads in a polymer depend on the Van der Waals and electrostatic forces.

The size of a dimer, the repeating unit in vWF chain, is estimated to be 80 nm in length [23]. Typical vWF multimers consist of 30 - 125 dimers [20]. The CNP of interest in our study has a diameter of 80 nm.

In the model-vWF, the vWF is composed of $N$ dimeric units and is represented as chain of $N$ beads. A CNP is modeled as an isolated bead suspended freely in fluid. The two-way interaction between fluid and any bead is modeled by the coupled lattice-Boltzmann and Langevin dynamics (LB-LD) method [21,22].

The dynamics of $i^{th}$ bead with position vector $\mathbf{r}_i$ is governed by the Langevin equation

$$\frac{d\mathbf{r}_i}{dt} = \mathbf{u}_\infty(\mathbf{r}_i) + \frac{1}{f_D}(\mathbf{F}_i^{ext} + \mathbf{F}_i^B(t)), \qquad (1)$$

where $f_D$ is the drag coefficient estimated by Stokes' law as $f_D = 6\pi\eta a$, $\eta$ is dynamic viscosity (for blood plasma $\eta =$ 1.2 cP), and $a$ is radius of a bead (for both dimer and nanoparticles, $a$ = 40 nm). The external, $\mathbf{F}_i^{ext}$, and Brownian, $\mathbf{F}_i^B$, forces are provided below. The characteristic time constant, $\tau = a^2 / \mu_0 k_B T$, is used as the time scale. Fluid velocity $\mathbf{u}_\infty(\mathbf{r}_i)$ is obtained by interpolation from discrete lattice sites via LB method.

The Brownian force has zero mean, and satisfies the fluctuation-dissipation theorem [24,25]:

$$\langle \mathbf{F}_i^B(t) \rangle = 0, \ \langle \mathbf{F}_i^B(t)\mathbf{F}_i^B(t') \rangle = 2k_B T f_D \delta_{ij}\delta(t-t')\mathbf{I}, \quad (2)$$

where $k_B$ is Boltzmann constant, $T$ is the temperature ($T$ = 298K), $\delta_{ij}$ is the Kronecker delta, $\delta(t-t')$ is the Dirac delta function, and $\mathbf{I}$ is the 2nd order unit tensor. The external force, $\mathbf{F}_i^{ext} = \nabla_{\mathbf{r}_i} U$, where $U$ is the total potential energy consists of bonded and non-bonded interactions. Bonded interactions result from biochemical bonds formed between two counterparts, like the force between two connecting beads in a vWF multimer. As bonds have limited extension even under strong extension, the bonds are reasonably modeled as finitely extendible nonlinear elastic (FENE) springs with potential, $U_{FENE}$ [14], given by

$$U_{FENE} = \sum_{i=1}^{N-1} -\frac{kR_0^2}{2}\ln(1-(\frac{r_{i,i+1}}{R_0})^2), \quad (3)$$

where $k$ is the spring constant, and $R_0$ denotes the maximum bond extension. Here, $k = 50k_B T/a^2$, $R_0$ =0.3 $a$ to ensure the polymer remains in the shape of a chain. Non-bonding interactions, on the other hand, act between atoms in the same molecule and those in other molecules, even at long range. Force fields divide non-bonding interactions into two: electrostatic interactions and Van der Waals interactions.

Electrostatic interaction arises due to the unequal distribution of charge in a molecule. The interaction between point charges is generally modelled by a Coulomb potential that acts along the line connecting the two charges. A particle's surface potential is difficult to obtain; various analytical schemes give different approximations. However, the zeta potential measured from electrophoretic mobility experiments is easily accessible, and perhaps the only way to characterize the charge magnitude of a colloid. It is assumed that a net positive charge is located at the center of the bead, and all negative ions locate on the center of the CNP as well. The magnitude of the charge is estimated using the experimentally measured zeta potential:

$$q_i = m \cdot (4\pi\varepsilon_r\varepsilon_0\zeta_i r_i) \quad (4)$$

where $r_i$ is the distance between particle center to slipping plane, approximately equal to radius of particle; $\varepsilon_r$ is relative vacuum dielectric constant of medium and $\varepsilon_0$ is vacuum permittivity. Due to the electrical double layer effect, there is an overestimation of net charge; therefore, an adjustment multiplier $m$ = 0.07 is included. $\zeta_i$ is the zeta potential of a charged particle. In simulations, $\zeta_i$ = -50 mV [26] for CNP and $\zeta_i$ = +400 mV for beads in chain (each one of two A1 patches has zeta potential of +150 to +250 mV [27,28]). Hence, we estimate Coulomb potential between two particles by:

$$U_C = \sum_{ij}\frac{1}{4\pi\varepsilon_r\varepsilon_0}\frac{q_i q_j}{r_{ij}} = \sum_{ij} m^2 \frac{4\pi\varepsilon_r\varepsilon_0\zeta_i\zeta_j r_i r_j}{r_{ij}}, \quad (5)$$

where $r_{ij}$ is the distance between two particles $i$ and $j$.

Van der Waal interactions, including dispersion, repulsion, and induction, are commonly considered as consisting of the all the interactions between atoms or molecules that are not covered by the electrostatic interaction [29]. The most common form used to model van der Waals interactions is the Lennard-Jones (LJ) potential:

$$U_{LJ} = \sigma k_B T \sum_{ij}((2a/r_{ij})^{12} - 2(2a/r_{ij})^6), \quad (6)$$

where σ determines the depth of the potential depending on molecule type. Here by choosing σ = 2.0, there is a clear unfolding transition at a well-defined critical shear rate around $\dot{\gamma}^*$ = 6000 s$^{-1}$ for vWF. This is consistent with existing experimental results, where fluorescent video-microscopy of a single vWF suspended in shear flow in a microfluidic channel [11] and BD simulations [10] show that the vWF remains globular up to a critical shear rate of ~5000 s$^{-1}$.

The LJ potential acts between all beads in the polymer chain and nanoparticles. At extremely short distance, the 12th power term in LJ potential will dominate any other potentials. This repulsive force accounts for the volume-exclusion effect.

**Results**

**vWF under high shear**

During a typical periodic cycle without CNP, thermal fluctuation turns a polymer orientation angle θ to negative (θ = -ε , where ε is a small positive number) and begins to collapse (Fig 1.1). The collapse along with the rotation process further decreases the orientation angle (θ << 0, Fig 1.2) and increases the vertical distance between the top and bottom of the polymer. The increase in vertical dimension accelerates the collapse. The acceleration reaches maximum at the globular shape when the vertical dimension reaches maximum. Since the cohesive force cannot overcome the shear force above the threshold shear rate, the globular shape cannot persist. After the transition of θ from negative to positive (from -π/2 to π/2, Fig 1.3) as it tumbles, the polymer elongates as θ decreases from θ >> 0 (Fig 1.4) to θ = +ε (Fig 1.5). When fully stretched, the orientation changes from θ = +ε to θ = -ε and begins another cycle. Unfolding starts with a local disturbance from fluid.

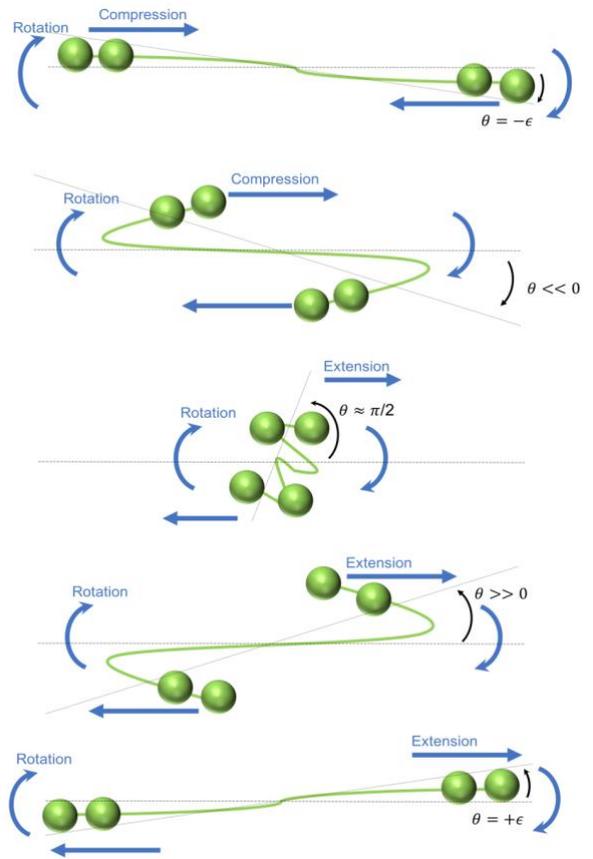

FIG. 1. Typical polymer dynamics under unbounded, steady shear flow above critical shear rate.

**vWF + CNP under high shear**

With CNP added, the vWF molecule behaves differently. There are two major differences in dynamics as seen in the shear plane. With CNP, the attachment of a CNP to a random location in the elongated chain introduces an *additional site* to initiate folding. This newly introduced disturbance by CNP gives a local impulse for folding. Second, if the CNP attaches to the polymer as it collapses and gets absorbed inside the globule, it can stabilize the coiled polymer by increasing the cohesive force. Thus, the new *complex* with CNP wrapped inside polymer can withstand shear stress, preventing the collapsed chain from elongating again. Fig 2 shows a side-to-side comparison of a chain interacting with or without CNP at the flow-gradient plane, under an unbounded, steady shear rate of 6500 s$^{-1}$. The elongation of normal vWF at 6500 s$^{-1}$ compared with the persistent collapse of vWF with CNP present.

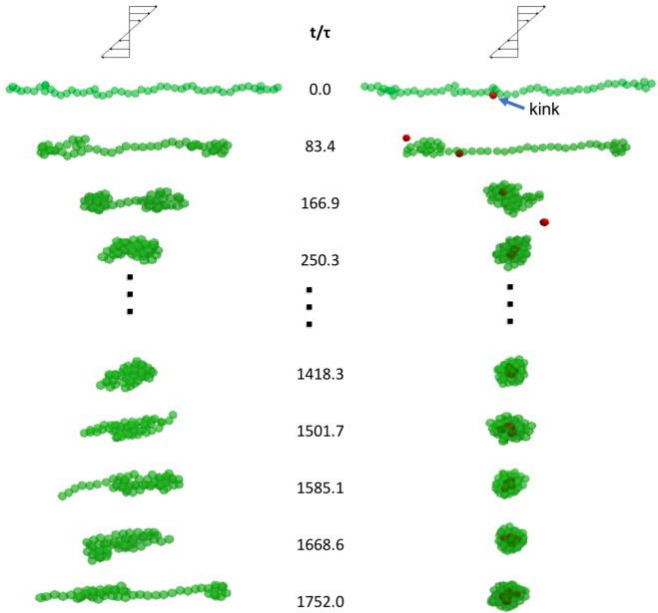

FIG. 2. Equal time interval snapshots of conformational dynamics of a single green vWF multimer without CNP (left) and with CNP (red dot, right) in simple shear flow at 6500 s$^{-1}$. Time t is scaled with the characteristic particle diffusion time constant τ. The CNP charge has zeta potential -50 mV. Each vWF dimer has +400 mV charge (two A1 domains), and there are 50 dimers in the model vWF. In this simulation, there are a few (<5) CNP available for vWF to observe interactions clearly. The electrostatic force of the CNP is strong enough to prevent elongation by vWF at a shear rate at 6500 s$^{-1}$.

Without CNP, typical cyclic dynamics of elongation, folding and tumbling are observed. The formation of a small protrusion can initiate unfolding that potentially leads to full elongation (normalized extension of polymer in flow direction $l^*$ > 0.7). In contrast, with CNP an additional local disturbance (kink) occurs around the position of CNP-polymer encounter. Shear-induced compression and rolling of vWF enhances globular formation. As some CNP get enclosed in the CNP-vWF composite, they stabilize the formed complex. The additional cohesive force is strong enough to prevent unfolding. Even with a small protrusion caused by thermal fluctuation, the tendency to re-coil dominates the tendency to stretch, resulting in a stable coiled shape, as shown in Fig 2 (right).

A subsequent question is if shear rate is further increased, will the coiled CNP-vWF composite elongate again? We further studied the combined effect of CNP charge and shear on polymer dynamics, quantitatively. For a single vWF in constant shear flow, it is shown that the critical shear rate $\dot{\gamma}^*$ scales as $\dot{\gamma}^* \sim N^{1/3}$, where $N$ is number of dimers in the polymer [10]. In the presence of CNP, this scaling is no longer complete. We explore the relation between $\dot{\gamma}^*$ and the CNP charge strength, designated as zeta potential, $\zeta$.

We explore the behavior of a single vWF chain under elevated shear rates at 13 000, 26 000, 39 000, 52 000 and 65 000 s$^{-1}$. This range of shear rate is of biological importance as most severe shear-induced platelet accumulation (SIPA) occurs in this range [30]. Under elevated shear rates (>13 000 s$^{-1}$), modeled-vWF elongates and does not assume a stable coiled conformation by itself, or with neutral nanoparticles (NP). Both simulations and experiments [31] show that neutrally charged NPs have negligible influence on the dynamics of vWF in flow.

Only negative CNP can significantly influence the conformational dynamics of the vWF. Denote zeta potential of CNP as $\zeta_1$ (< 0) and zeta potential of large patches of positive charge at vWF-A1 domains as $\zeta_2$ (> 0). The magnitude of $\zeta_1$ is referred to as $\zeta = |\zeta_1|$ for the rest of the article. $\zeta_2$ = +400 mV [27] and viewed as a constant in the simulations. In a series of simulations, $\zeta$ gradually increases from 0 mV up to a point where stable CNP-vWF composite is formed.

Besides CNP charge, another parameter that could influence critical shear rate is the vWF size or the number of beads ($N_{vWF}$). It is estimated that there are 30 - 125 dimers in a typical plasma vWF [20]. Thus 50, 100 and 200 beads are considered in the model-vWF.

We set the number of CNP to total beads in vWF to be 2:1 in order to allow an abundance of CNP available to interact with vWF across all simulation cases, and to ensure that a CNP-vWF composite is formed at coiled equilibrium state. An example of the composite is shown in Fig 3 for $N_{vWF}$ = 50.

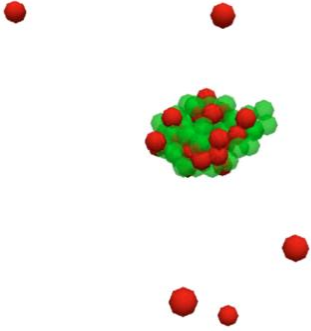

FIG. 3. An example of a stable globular CNP-vWF composite. Red beads are CNP, the green chain is vWF. Since CNP are abundant, the number of CNP in the coiled CNP-vWF composite, is limited by the size of vWF. In this case, $N_{vWF}$ = 50, $N_{CNP} \approx 53$. After sufficient simulation time, although CNP continues to attach and detach, $N_{CNP}$ almost remains unchanged over time. This coiled state is defined as a *dynamic equilibrium state* for the composite.

For a vWF of given size under constant shear rate (fixed $N_{vWF}$ and $\dot{\gamma}$), the critical zeta potential is found by the method described here. CNP charge is initially set to zero ($\zeta$ = 0 mV). Subsequently, the charge assigned to CNP is increased in a stepwise fashion from 0 mV to 48 mV, with 1 mV increments. Each time $\dot{\gamma}$ takes a new higher magnitude, at this pair of $\dot{\gamma}$ and $\zeta$, we expect the CNP-vWF system to reach a new dynamic equilibrium state; and that the asymptotic time-averaged normalized vWF length, $l^*$, tends to decrease due to the greater CNP effect. The charge density of CNP is increased by increasing $\zeta$ value to a point where periodic oscillation behavior of the polymer (e.g., Fig 1) ceases, and the polymer rolls back to a stable globular shape. This critical transition state corresponds to a critical charge (characterized by a critical zeta potential, $\zeta^*$) of CNP under that shear rate, i.e. $\zeta^* = \zeta^*(\dot{\gamma})$. The process to find $\zeta^*(\dot{\gamma})$ is illustrated in Fig 4. For example, at shear rate $\dot{\gamma}$ = 52 000 s$^{-1}$, $N_{vWF}$ = 100, a series of tests of charges ranging from 0 to 34 mV are performed. Time evolution of normalized vWF extension $l^*$ at $\zeta$ = 26, 30 and 34 mV are selected and presented in this figure. For this high shear rate of 52 000 s$^{-1}$, the transition of vWF dynamics occurs for a CNP zeta potential of ~ 30mV. Or conversely, CNP of -30 mV will keep vWF in the globular state up to shear rate 52 000 s$^{-1}$.

Below a critical charge magnitude, vWF still oscillates and can be elongated under shear (Fig 3(a)). Above the critical charge $\zeta^*$ (Fig 3(b)), a stable globular CNP-vWF composite is observed after initial transient period (t/τ = 0 ~ 50). Continuing to add charge to the CNP will keep the globular CNP-vWF composite stable under the same flow condition (Fig 3(c)). Quantitatively, we can define $\zeta^*$ as when post-transient average normalized extension $l^*$ is always less than 0.20. In the case of Fig 4, for instance, $\zeta^*$ is 28 mV.

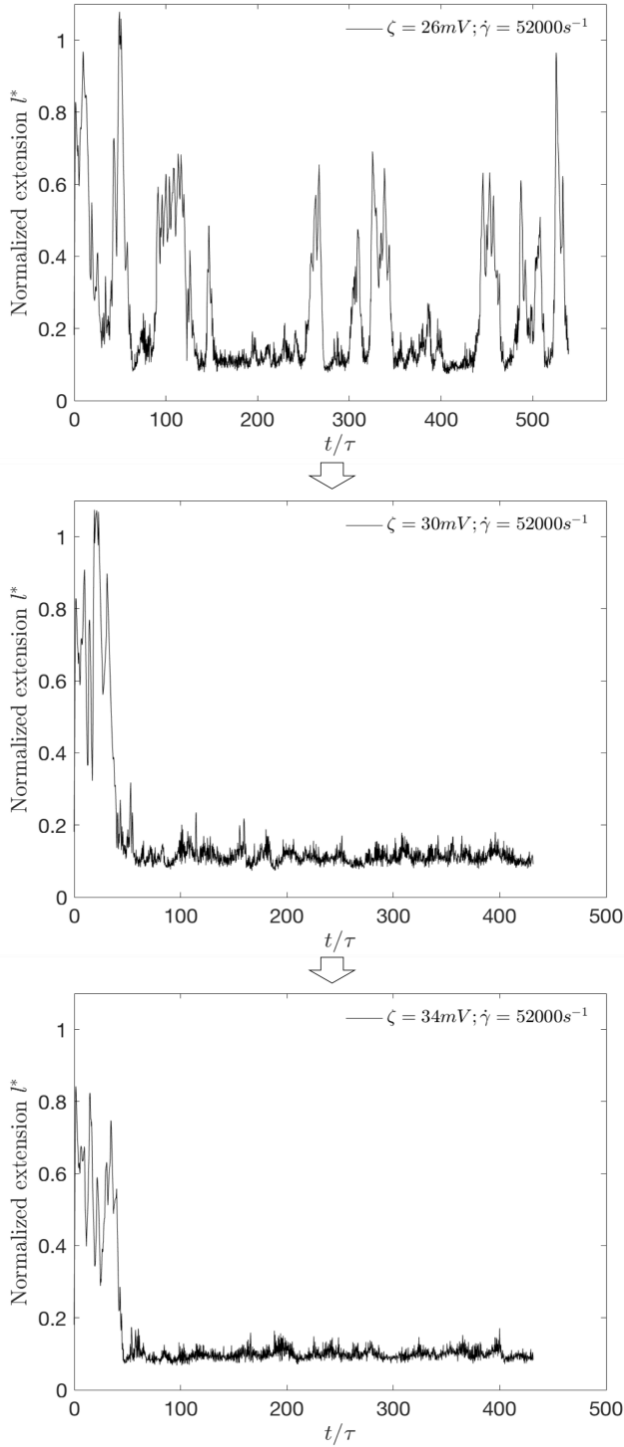

FIG. 4. Example of exploration of critical charge $\zeta^*$ by increasing magnitude of $\zeta$ starting from 0 mV. Shear rate $\dot{\gamma}$ = 52 000 s$^{-1}$, $N_{vWF}$ = 100. Three explorations at $\zeta$ = 26, 30, and 34 mV are presented. In each case, at a constant charge, the simulation lasts long enough for CNP and vWF interaction to reach an equilibrium. From $\zeta$ =0 mV, $\zeta$ gradually increases to a point where a stable globular complex is first formed. Phases of polymer dynamics change dramatically in between 26 mV (Fig 4(a)) and 30 mV (Fig 4(b)). Beyond 30 mV, when $\zeta$ is further elevated, the CNP-vWF composite remains in the coiled, globular shape (Fig 4(c) after initial transit, corresponds to Fig 3). This transition in states indicates a critical charge (for $\dot{\gamma}$ =52 000 s$^{-1}$) exists within range $\zeta$ = [26 mV, 30 mV]. Refining the search further determines $\zeta^*$ = 28 mV.

A similar search for $\zeta^*$ is performed for 5 shear rates with $N_{vWF}$ = 50, 100, and 200. The results are summarized in Fig 5. For a given vWF length, $\zeta^*$ scales linearly with shear rate $\dot{\gamma}$. Also, $\zeta^* \sim N_{vWF}^{-2/5}$ as shown in Fig 6.

At the final coiled equilibrium state, the resulting composite consists of both vWF and CNP (Fig 3). The dimension of the composite is determined by $(N_{vWF} + N_{CNP})$ rather than $N_{vWF}$ alone. So, it is more relevant to find the scaling relationship between $\zeta^*$ and $(N_{vWF} + N_{CNP})$. However, $N_{CNP}$ also depends on $N_{vWF}$ and charge, as: 1) $N_{vWF}$ limits the number of CNP that can be attached, and 2) charge impacts the strength of cohesive force in the CNP-vWF composite. Here, we define a new parameter $\beta = (N_{vWF} + N_{CNP})/N_{vWF}$, as the ratio of total size of composite normalized by the size of polymer itself. $\beta$ is dependent on $N_{vWF}$ in a nonlinear relationship, but less dependent on shear rate, as shown in Fig 7. When $\beta(N_{vWF})$ instead of $N_{vWF}$ is used in scaling, it turns out that $\zeta^* \sim \beta^{-3}$ (Fig 8).

Thus, a scaling relation is derived with regards to a critical equilibrium state, as $\zeta^* \sim \dot{\gamma}\beta^{-3}$. At this state, the CNP-vWF composite is in a globular shape and there is an almost constant ratio $\beta$ between composite size and vWF size. The shear stress imposed by fluid flow is balanced by the cohesive force within the composite. The charge is just strong enough to hold together the refolded vWF, and the shear rate $\dot{\gamma}$ is just weak enough not to elongate the coiled vWF polymer. In other words, the equilibrium state of coiled composite can be broken by either increasing shear rate $\dot{\gamma}$ or decreasing charge $\zeta$.

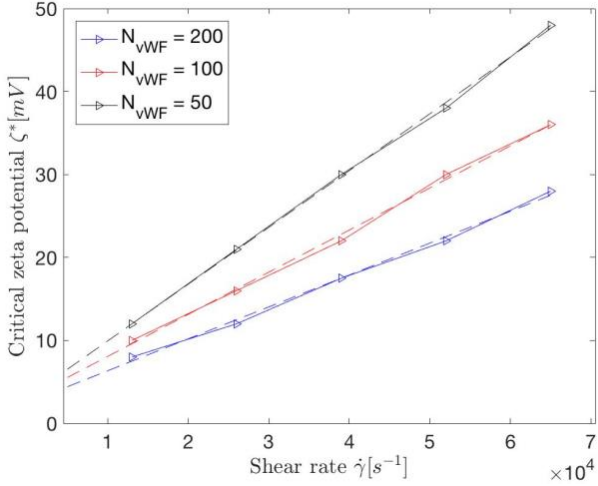

FIG. 5. Critical charge as a function of shear rate for vWF of size 50, 100, and 200. For constant $N_{vWF}$, critical charge scales linearly with shear rate. Dashed lines are linear interpolation of data points (triangles). Note that the lines extrapolate to ($\dot{\gamma}$ = 6500 s$^{-1}$, $\zeta$ = 5 mV) rather than $\zeta$ = 0 mV. This indicates that weakly charged CNP ($\zeta$ < 5 mV) has no significant effect, just like neutral NP.

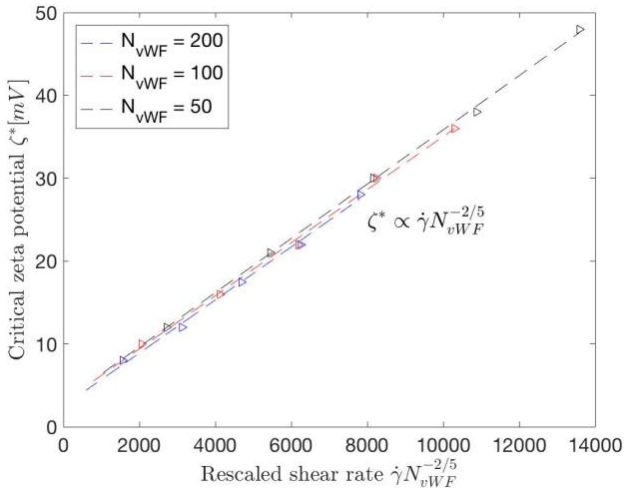

FIG. 6. Critical charge as a function of shear rate rescaled by $N_{vWF}^{-2/5}$ for vWF of size 50, 100, and 200. $\zeta^* \sim \dot{\gamma} N_{vWF}^{-2/5}$.

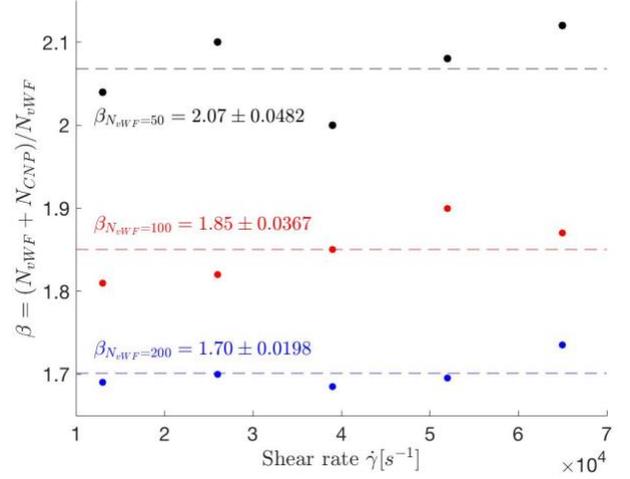

FIG. 7. Ratio of total size of composite normalized by the size of polymer, $\beta = (N_{vWF} + N_{CNP})/N_{vWF}$ versus shear rate, with vWF size 50, 100, and 200, respectively. $\beta$ is strongly dependent on $N_{vWF}$ but not on shear rate. For a given vWF size, $\beta$ has no discernible dependency on shear rate. As vWF size increases, the mean of $\beta$ (horizontal dash line) decreases, so does the deviation from the mean.

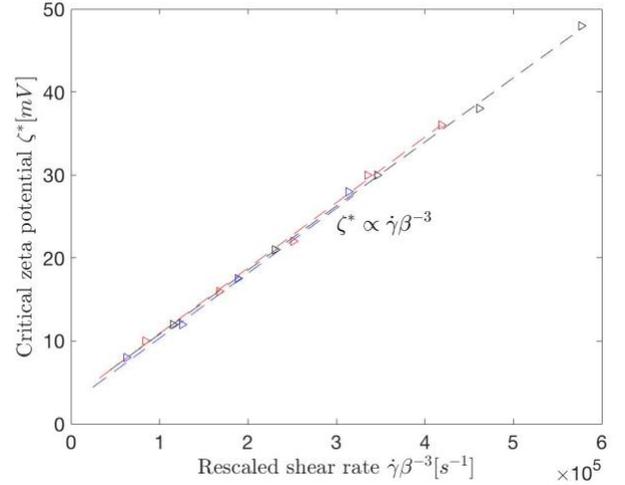

FIG. 8. Critical charge as a function of shear rate rescaled by $\beta^{-3}$ for vWF of size 50, 100, and 200. $\zeta^* \sim \beta^{-3}$. The dependence of $\beta$ on shear rate and vWF size is shown in Fig 7.

In conclusion, by observing the interactions of CNP on the dynamics of polymer under shear flow, we provide theoretical evidence for the shift of critical shear rate to much greater values after introducing CNP. If a polymer undergoes a conformational dynamics as represented in Fig. 1, adding CNP with opposite charge (-) to the polymer's surface charge (+) could significantly alter the conformational dynamics of the polymer. As shown here, addition of CNP can elevate the critical shear rate and change the scaling relation with respect to polymer length. We show that the critical zeta potential of CNP, $\zeta^*$, scales linearly with shear rate, and scales cubically with $\beta$, that is $\zeta^* \sim \dot{\gamma}\beta^{-3}$.

These results have relevance to hemostasis, as high-shear induced elongation of proteins, such as vWF could be detrimental as it is central to arterial thrombosis, leading to myocardial infarction or stroke. An elevation in the critical shear rate may lower the possibility of the shear induced thrombosis. Actively controlling the critical shear rate where abrupt length change occurs has potential to control polymer system dynamics.

**References**

[1] P. G. De Gennes, J. Chem. Phys. **60** (1974).
[2] D. E. Smith, H. P. Babcock, and S. Chu, Science **283** (1999).
[3] R. G. Larson, J. Rheology (N.Y.) **49** (2005).
[4] P. E. Rouse, The Journal of Chemical Physics **21**, 1272 (1953).
[5] B. H. Zimm, The Journal of Chemical Physics **24**, 269 (1956).
[6] A. Link and J. Springer, Macromolecules **26**, 464 (1993).
[7] E. C. Lee, M. J. Solomon, and S. J. Muller, Macromolecules **30**, 7313 (1997).
[8] D. E. Smith, H. P. Babcock, and S. Chu, Science **283**, 1724 (1999).
[9] R. G. Larson, H. Hu, D. E. Smith, and S. Chu, Journal of Rheology (1978-present) **43**, 267 (1999).
[10] A. Alexander-Katz, M. F. Schneider, S. W. Schneider, A. Wixforth, and R. R. Netz, Phys Rev Lett **97**, 138101 (2006).
[11] S. W. Schneider, S. Nuschele, A. Wixforth, C. Gorzelanny, A. Alexander-Katz, R. R. Netz, and M. F. Schneider, Proc Natl Acad Sci U S A **104**, 7899 (2007).
[12] C. A. Siedlecki, B. J. Lestini, K. K. Kottke-Marchant, S. J. Eppell, D. L. Wilson, and R. E. Marchant, Blood **88**, 2939 (1996).
[13] Z. M. Ruggeri, Best Pract Res Clin Haematol **14**, 257 (2001).
[14] K. Kremer and G. S. Grest, The Journal of Chemical Physics **92**, 5057 (1990).
[15] P. Ahlrichs and B. Dunweg, Journal of Chemical Physics **111**, 8225 (1999).
[16] P. S. Doyle and P. T. Underhill, Handbook of Materials Modeling, 2619 (2005).
[17] A. Alexander-Katz and R. R. Netz, Macromolecules **41**, 3363 (2008).
[18] H. Fu, Y. Jiang, D. Yang, F. Scheiflinger, W. P. Wong, and T. A. Springer, Nature communications **8**, 324 (2017).
[19] M. T. Griffin, Y. Zhu, Z. Liu, C. K. Aidun, and D. N. Ku, BioMicrofluidics (2018).
[20] T. A. Springer, J Thromb Haemost **9 Suppl 1**, 130 (2011).
[21] Z. Liu, Y. Zhu, R. R. Rao, J. R. Clausen, and C. K. Aidun, arXiv preprint arXiv:1801.02299 (2018).
[22] Z. Liu, Y. Zhu, R. R. Rao, J. R. Clausen, and C. K. Aidun, Computers & Fluids (2018).
[23] I. Singh, H. Shankaran, M. E. Beauharnois, Z. Xiao, P. Alexandridis, and S. Neelamegham, J Biol Chem **281**, 38266 (2006).
[24] R. Kubo, M. Toda, and N. Hashitsume, *Statistical physics II: nonequilibrium statistical mechanics* (Springer Science & Business Media, 2012), Vol. 31.
[25] R. G. Larson, Journal of Rheology **49**, 1 (2005).
[26] S. K. Kloet, A. P. Walczak, J. Louisse, H. H. van den Berg, H. Bouwmeester, P. Tromp, R. G. Fokkink, and I. M. Rietjens, Toxicol In Vitro **29**, 1701 (2015).
[27] R. A. Romijn, PhD thesis, 2003.
[28] E. G. Huizinga, S. Tsuji, R. A. Romijn, M. E. Schiphorst, P. G. de Groot, J. J. Sixma, and P. Gros, Science **297**, 1176 (2002).
[29] U. Burkert and N. Allinger, Washington, DC (1982).
[30] L. D. Casa, D. H. Deaton, and D. N. Ku, J Vasc Surg **61**, 1068 (2015).
[31] A. Nemmar, M. F. Hoylaerts, P. H. Hoet, D. Dinsdale, T. Smith, H. Xu, J. Vermylen, and B. Nemery, Am J Respir Crit Care Med **166**, 998 (2002).